\documentclass{article}

\PassOptionsToPackage{numbers, compress}{natbib}

\usepackage[preprint]{neurips_2024}

\usepackage[utf8]{inputenc} 
\usepackage[T1]{fontenc}    
\usepackage{hyperref}       
\usepackage{url}            
\usepackage{booktabs}       
\usepackage{amsfonts}       
\usepackage{nicefrac}       
\usepackage{microtype}      

\usepackage{cleveref}
\usepackage{graphicx}
\usepackage{wrapfig}
\usepackage{booktabs}
\usepackage[table,xcdraw]{xcolor}
\usepackage{svg}

\title{A Foundation Model for the\\Solar Dynamics Observatory}

\author{
James Walsh\\University of Cambridge
\And
Daniel G. Gass\\University of Central Lancashire
\And
Raul Ramos Pollan\\Universidad Industrial de Santander
\And
Paul J. Wright\\Dublin Institute for Advanced Studies
\And
Richard Galvez\\Pure Storage
\And
Noah Kasmanoff\\AE Studio
\And
Jason Naradowsky\\University of Tokyo
\And
Anne Spalding\\Trillium Technologies Inc
\And
James Parr\\Trillium Technologies Inc
\And
Atılım Güneş Baydin\\University of Oxford
}

\begin{document}

\maketitle

\begin{abstract}
SDO-FM is a foundation model using data from NASA's Solar Dynamics Observatory (SDO) spacecraft; integrating three separate instruments to encapsulate the Sun’s complex physical interactions into a multi-modal embedding space. This model can be used to streamline scientific investigations involving SDO by making the enormous datasets more computationally accessible for heliophysics research and enable investigations that require instrument fusion. We discuss four key components: an ingestion pipeline to create machine learning ready datasets, the model architecture and training approach, resultant embeddings \& fine-tunable models, and finally downstream fine-tuned applications. A key component of this effort has been to include subject matter specialists at each stage of development; reviewing the scientific value and providing guidance for model architecture, dataset, and training paradigm decisions. This paper marks release of our pretrained models and embedding datasets, available to the community on Hugging Face and \href{https://sdofm.org}{\textit{sdofm.org}}.
\end{abstract}

\section{Introduction\label{sec:intro}}

NASA's Solar Dynamics Observatory (SDO) \cite{2012SoPh275207S} is a space based heliophysics laboratory, capturing details of the Sun's atmosphere and photosphere across a wide variety of spatial and temporal scales, wavelengths and levels of activity across the solar cycle. It was launched in 2010 as part of NASA's Living With a Star Program and has been instrumental in advancing the fields of observational heliophysics. It has captured the entirety of Solar Cycle 24 (2010 - 2020), and is currently observing the peak of Solar Cycle 25 (2021 - ongoing). The SDO consists of a suite of three instruments; the Helioseismic and Magnetic Imager (HMI) \cite{2012SoPh275207S} which captures information on the Sun's magnetic field at the photosphere, the Atmospheric Imaging Assembly (AIA) \cite{2012SoPh27517L} which captures ultraviolet and soft X-ray radiation emitted from the Sun's atmosphere in ten channels from 10-100 nm, and the Extreme-ultraviolet Variability Experiment (EVE) \cite{2012SoPh..275..115W}, which produces spectra for extreme ultraviolet irradiance reaching Earth's atmosphere. This suite of instruments represents a large dataset suitable for use in training a foundation model, as well as a dataset with direct relevance to ongoing scientific research. The unique challenges and properties of the SDO dataset are outlined in Section 1.1.

The overall process for building foundation model for SDO, known as SDO-FM, is composed of four stages; (1) data preparation, (2) large foundation model (FM) training, (3) embedding extraction, and (4) fine-tuning or direct embedding usage for scientific validation cases. Collectively we denote the data preparation as effort completed under SDOML \cite{galvez2019machine}, a machine-learning dataset of SDO. Our models are based upon autoencoders, with training conducted under the objective of image reconstruction over the period beginning from satellite launch in 2010 to 2023. Once these models are trained, a compressed representation dataset is created from the embeddings by a full-pass over the encoder. The compressed representations are called direct embeddings and provide a helpful result as a set of available SDO features at around two-thousandths (0.002) the original size. Lastly, the direct embeddings as well as standard model fine-tuning, are used to conduct scientific validation through a validation harness which is used to check our results against past ML-based heliophysics approaches and to compare their computational expense.

\subsection{Input Data}

To properly represent the native SDO dataset in a format which is compatible with conventional machine learning architectures, the SDOML dataset \cite{galvez2019machine} was used. This dataset is available at {\url{http://sdoml.org}}, and is comprised of 14 years of SDO data from all three instruments (HMI, AIA, and EVE). The high resolution and time cadence of AIA (4096 x 4096 pixels, 12 second cadence) and HMI (4096 x 4096) results in a database greater than 22 Pb in size on disk, comprised of multiple domains of image, magnetic flux, and spectra, as well as multiple image wavelengths.

To reduce the 14 years of data, SDOML v2.0 is offered at a fixed resolution and cadence (512x512 pixels every 12-minutes), with standardization and corrections. This includes transformations from Level-1 to 1.5 \cite{Barnes2020}. The size of the reduced dataset on disk of the 14-year period is roughly 10Tb, a more manageable state compared to the 22Pb source.

\subsection{Scientific Direction of SDO-FM}

The objective of this foundation model for heliophysics is to encode a representation of the phenomena observed by the SDO spacecraft. This includes key aspects of solar physics, such as the relationship between the solar magnetic field, eruptions (flaring), and extreme ultraviolet irradiance, as well as dynamics such as differential rotation, convection \& magnetic processes, and observational factors such as instrumental degradation. Once physical phenomena are encoded in the embedding space, researchers can then develop fine-tuned downstream applications to explore how these phenomena impact the wider solar system - known as the Heliosphere. To achieve this ambitious goal, we begin with encoding known phenomena through a latent representation discussed further in \cref{sec:model_choice}. The model is then validated by adapting the embeddings and comparing outcomes against published known results using classical machine learning methods. The classical comparisons were selected for applicability to the compute constraint problem: predicting the Earth's thermospheric density (F10.7), reconstructing missing channels, autocalibration of the AIA instrument, and the virtualization of the broken MEGS-A instrument, detailed in \cref{sec:val_cases}.

\section{Related Work}
The idea for this approach came from previous efforts to encode the solar disk \cite{brown2021learning} and employ those embeddings to open heliophysics challenges, such as thermospheric density prediction \cite{malik2023high}. It was evident from these investigations that there was merit to including SDO EUV data to improve those domain objectives, however these initial efforts were impeded by the high computational cost of training with large-scale image datasets, prompting the need for a foundational community asset that could be adapted to multiple heliophysics and space weather use-cases.  

The precedent for employment of the MAE came from the proposal of MAEs as spatiotemporal learners \cite{feichtenhofer2022masked} and the NASA IMPACT-IBM progress on a generalizable geospatial foundation model on remote sensing imagery \cite{jakubik2023foundationmodelsgeneralistgeospatial}.

\begin{figure}[h]
\includegraphics[width=\textwidth]{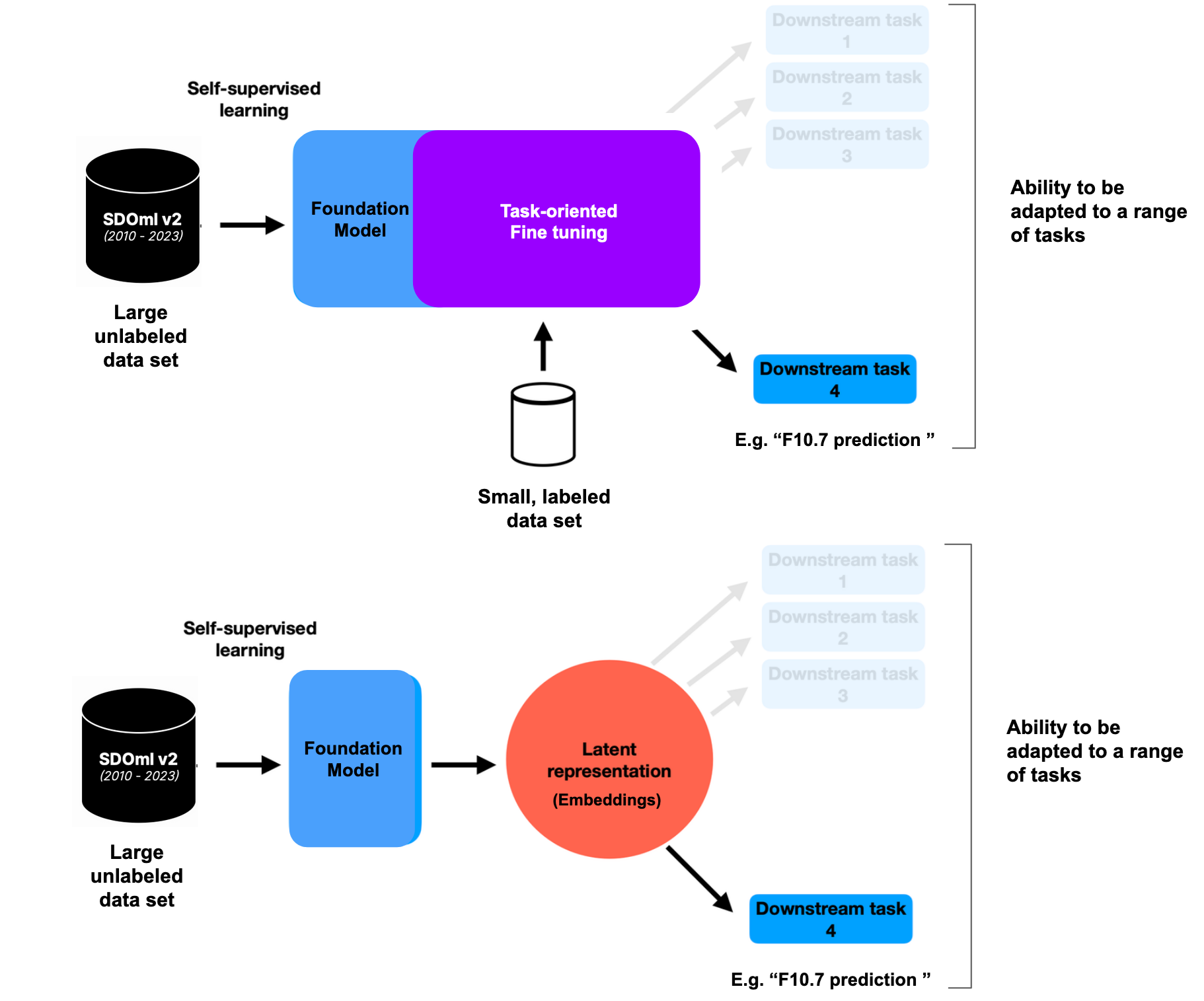}
\centering
\caption{Two methods of using the pre-trained backbone, directly with an adaptor for fine-tuning \cite{arch_diags}, or with a new model consuming the generated latent representation directly.}
\end{figure}

\section{Method\label{sec:method}}
Our foundation model is composed of a \textit{backbone}, optional \textit{neck}, and \textit{head}. We define the backbone as the model initially trained on the reconstruction task, the neck as the converter between backbone and head, the former then selected for the downstream application (or validation task). We implement two model families as backbones, one stemming from a Nouveau Variational Autoencoder (NVAE) \cite{vahdat2020nvae}, the other from a MAE \cite{he2022masked}. They are both adapted to better accommodate our scientific dataset and for intermediate export of their latent spaces in the form of ``embeddings.'' We additionally evaluated various feature engineering options regarding how to manage the solar disk, the most effective included a simple look up from Stonyhurst coordinates, a heliographic coordinate system for a fixed observer on Earth (suitable given the geosynchronous orbit), to pixel space.

\subsection{Model choice\label{sec:model_choice}}

Model selection was initially determined by the ability to capture solar phenomena, guided by applicability to SDO imagery and the ease of access to the embeddings in the latent space. The autoencoder architecture was selected for the backbone for ease of embedding construction and extraction. By design, autoencoders create a lower-dimensional representation during the encoding process. Other requirements included engineering efficiencies, such as ability to mask the solar limb for on-disk experiments, and cheaply bias by importance sampling for areas of interest (e.g. active regions).

\begin{figure}[h]
\includegraphics[width=\textwidth]{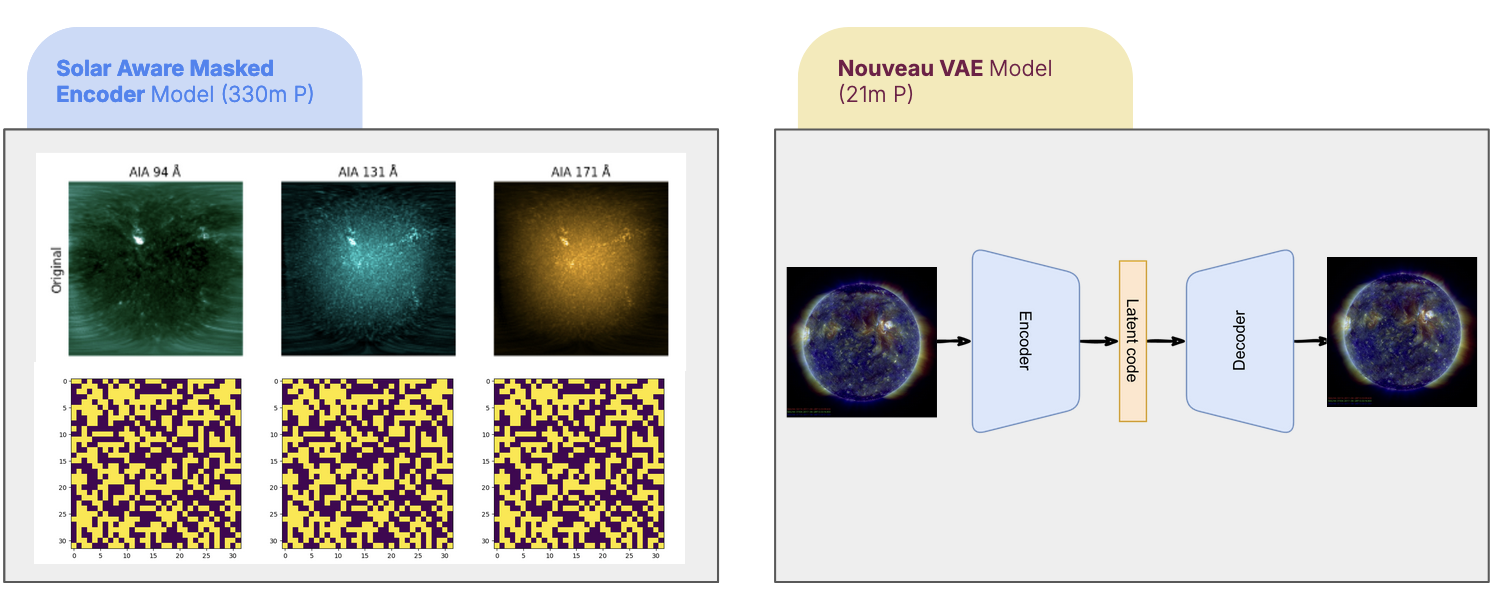}
\centering
\caption{Visualizations of the \textsc{samae} (left) and Nouveau-VAE (right), where the \textsc{samae} input has been transformed for full coverage, and the original Nouveau-VAE code base expanded to enable extraction of the latent representation.\label{fig:2}}
\end{figure}

\subsubsection{Solar-aware Masked Autoencoder}
Masked Autoencoders (MAEs) learn to be capable at reconstructing images with random components removed \cite{he2021maskedautoencodersscalablevision}. The approach follows the standard ViT-patchification common to transformer computer vision approaches for deconstruction of the image that the attention mechanism can learn between. The source of this ``powerful expressivity'' is attributed to ``rich hidden representation'' \cite{cao2022understandmaskedautoencoders}. This is particularly of interest in our scenario, as we seek to learn which components of solar imagery are of value for our scientific validation cases. This model was expanded to increase suitability for temporal information for remote sensing tasks \cite{Prithvi-100M-preprint}. We have continued to iterate, adding ``solar-awareness'' by including the ability to process the nine wavelengths of interest to us via the Atmospheric Imaging Assembly, efficiencies for processing the solar disk, and the ability to optionally bias the model towards learning active regions of scientific interest.

\subsubsection{Nouveau-VAE}
The Nouveau Variational Autoencoder (\textsc{NVAE}) is a deep hierarchical VAE created for image generation. Like the MAE, it is able to create a rich latent space using depth-wise separable convolutions and batch normalization. The \textsc{NVidia} team's codebase was modified to permit access to the hierarchical structure to successfully extract embeddings \cref{fig:2}.

\subsection{Scientific Validation Cases\label{sec:val_cases}}

\paragraph{Predict F10.7}
This index is a proxy for solar irradiance, which can be measured from the ground, as this frequency is not absorbed by the atmosphere. Can we achieve good agreement with ground measurements? There is limited scientific value in this prediction of a proxy measure such as F10.7, however this simple task clearly indicates learned capacity in a single result.

\paragraph{Virtual EVE}

In 2014, an instrument malfunction resulted in the loss of the MEGS-A module of SDO/EVE. With four years of overlapping data, \cite{2019SciA....5.6548S, indaco2024virtualevedeeplearning} used a hybrid CNN/linear regression model to successfully demonstrate the capability of machine learning methods to estimate missing EUV irradiance measurements from MEGS-A (and the degraded MEGS-B components of the EVE instrument). This validation task employs the embeddings constructed from AIA to understand the contributions from solar features on the EUV spectra, as the mapping between instruments exists due to the narrow-band images (SDO/AIA) and sun-as-a-star spectra (SDO/EVE) observing the same plasma distribution. A linear model accounts for a large portion of the relationship, while a CNN is used to correct for outlier events such as solar flares. There are known concerns regarding the model's performance post-2020, as AIA instrument performance deviates further from the 2014 baseline. Some of these issues can be addressed by incorporating other sources of irradiance, such as data from sounding rockets, for training over longer periods, although these are sparse. Importantly, this outperforms a physics-based inversion approach \cite{2019zndo...2587015W}. 

\begin{figure}[h]
    \centering
    \hspace*{2.8cm}
    \includesvg[width=1\textwidth]{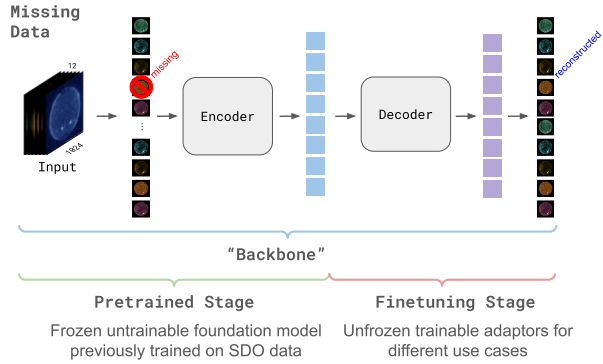}
    \caption{Missing/corrupt data reconstruction process.}
    \label{fig:arch_missing_data}
\end{figure}

\paragraph{Missing Channel Reconstruction} 
The reconstruction of missing extreme ultraviolet (EUV) images from wavelength images is a crucial task given the often low or unusable quality of image data frames from the Solar Dynamics Observatory (SDO). Currently, there is no effective method to recover these missing steps. However, the foundation model is capable of reconstructing individual frames by leveraging contextual information available in other wavelength channels. This approach allows for interpolation to provide a best-guess estimate of missing data at any arbitrary time step.

As with the Virtual EVE project, and differential emission measure analysis \cite{2019zndo...2587015W}, the overlapping temperature range covered by different SDO/AIA wavelength channels allows for the temperature distribution of the underlying plasma to be reconstructed, may enable the inference of properties of different temperature ranges.

The overlapping range covered by different wavelength channels may enable the inference of properties of different temperature ranges. This overlap can be used within a machine learning model to produce an estimation to replace data that is either missing, corrupted, or otherwise unusable. Our objective is to develop a more robust model that operates with higher computational efficiency while producing results comparable to the current SOTA. Special attention is given to the model's ability to capture non-linear relationships or rare events, such as intensity values in flaring regions.

There are several uncertainties inherent in this process. Some channels may be more readily recreated than others due to the physical assumptions that channels in the middle of the temperature/wavelength ranges will have the most overlap with other channels, potentially yielding better results. However, this overlap might not always correspond to the actual missing data in the SDO. Addressing these uncertainties requires an understanding of the shortfalls to determine the appropriateness of this reconstruction technique in different scenarios.

\begin{figure}[h]
    \centering
    \includesvg[width=1\textwidth]{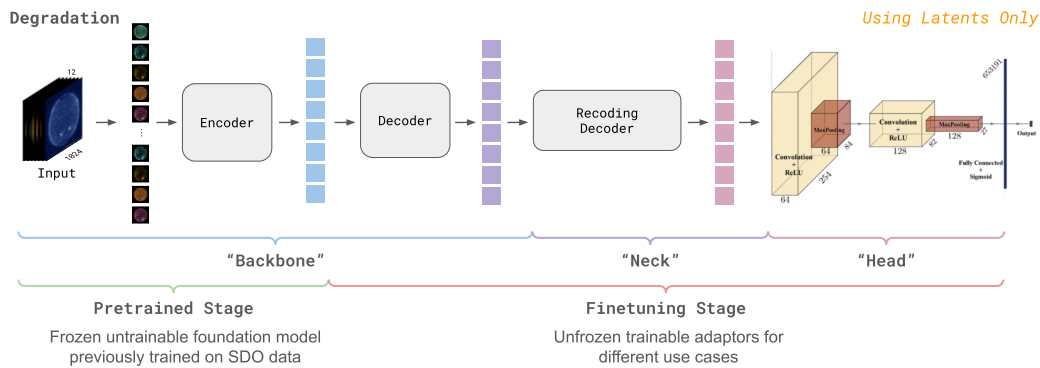}
    \caption{Instrument degradation prediction, head architecture reproduced with permission, \cite{2021AA648A53D}.}
    \label{fig:arch_missing_data}
\end{figure}

\paragraph{Autocalibration}

The SDO/AIA EUV channels exhibit degradation due to exposure to the same emissions they are intended to measure. This degradation results in apparent dimming over time across multiple EUV channels with unique characteristics. This poses challenges for long-term studies, as degradation trends within the dataset need to be corrected. Until 2014, SDO utilized EVE to correct this degradation. As discussed, a malfunction of SDO/EVE resulted in the loss of the MEGS-A component, and calibration is currently performed by sounding rocket flights. In response to this, \cite{2021AA648A53D} used a CNN to reconstruct the Atmospheric Imaging Assembly (AIA) multi-channel degradation curves.

Data requirements for this study include the SDOML data from AIA as well as older correction tables. The sampling requirement is minimal, with data being required once per day or even less frequently. Traditional SOTA methods, such as those performed by the Lockheed Martin Solar and Astrophysics Laboratory (LMSAL), involve calibration using sounding rocket flights. These methods, while accurate, are expensive and technically demanding. Our goal is to reproduce the results from \cite{2021AA648A53D} with greater efficiency in terms of data required and computational resources. This efficiency is evaluated through an examination of the resultant images compared to those produced by SOTA calibration pipelines, alongside intensity histograms, data spike analysis, and other metrics.

\section{Results\label{sec:results}}
Overall, our model families were evaluated for their backbone reconstruction task and against our four scientific validation cases. In all but the autocalibration task they reached the same level of accuracy or surpassed their classical counterparts in a fraction of the required time. In the autocalibration case, the direct embedding approach was able to match, but took additional training time. 

\subsection{Reconstruction}
\begin{figure}[h]
\includegraphics[width=\textwidth]{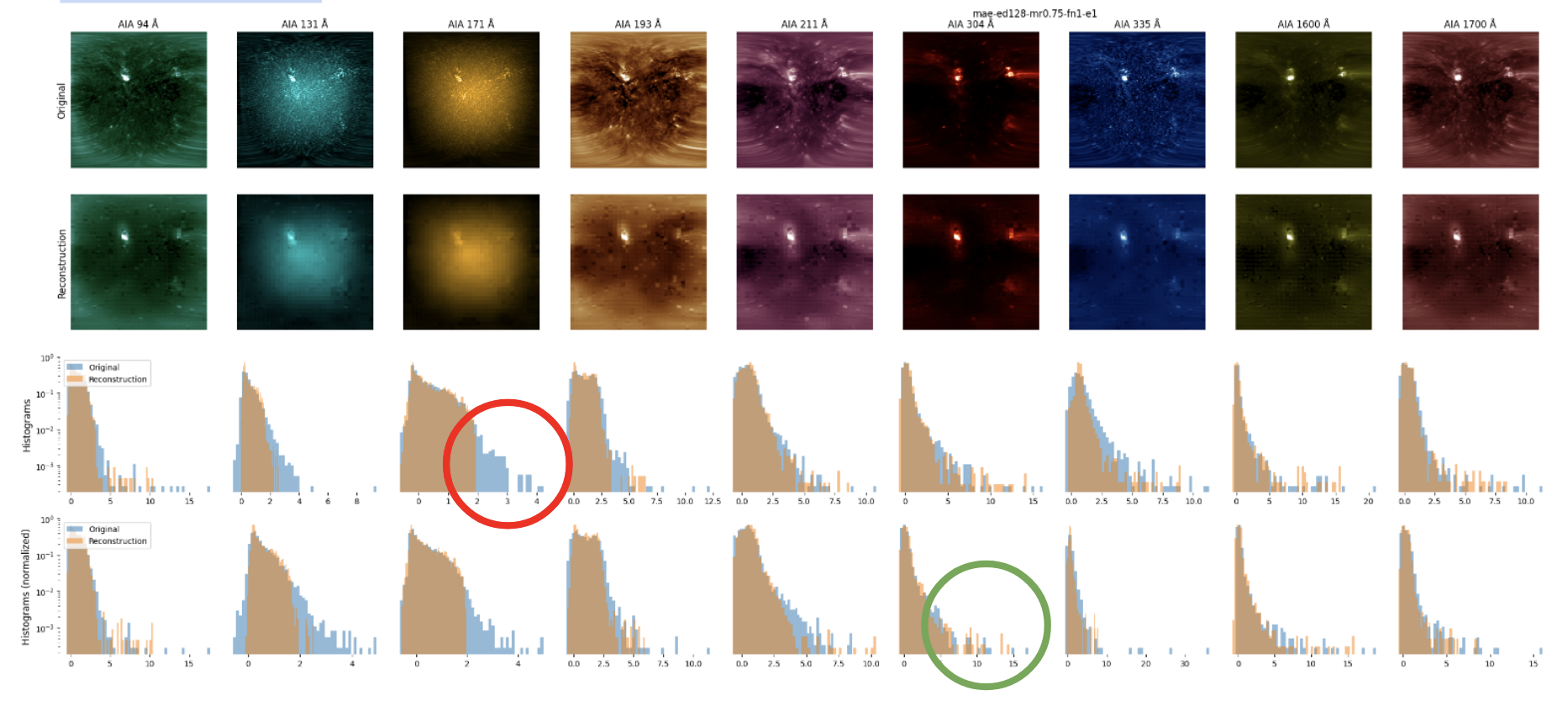}
\centering
\caption{\textsc{samae} reconstruction with disk transform, notice how in the circled areas peaks of some wavelengths are not captured well.\label{fig:recon_samae}}
\end{figure}

\begin{figure}[h]
\includegraphics[width=\textwidth]{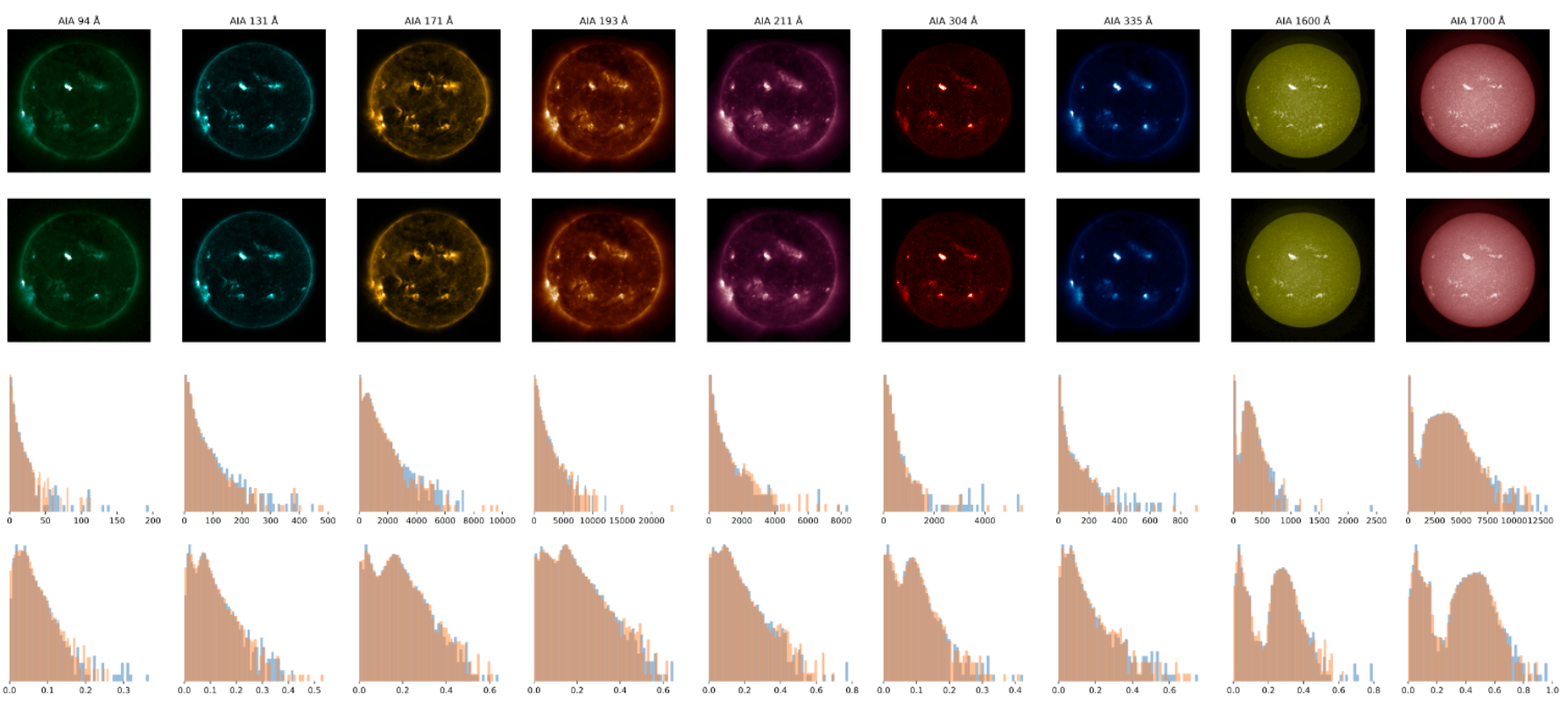}
\centering
\caption{Nouvaeu-VAE reconstruction, with better performance.\label{fig:recon_nvae}}
\end{figure}

Loss for the reconstruction task is measured by pixel RMSE within the solar disk. \textsc{samae} results presented in \cref{fig:recon_samae} indicate a clear ability to reconstruct most wavelengths under a \textit{small} embedding dimension (128) and within a \textit{short} number of training epochs (10). Interestingly this model struggles to reconstruct 131 \& 171\r{A}, which is likely due to a normalization error we're still investigating. The Nouveau-VAE model on raw pixel intensity performs better, even when including the solar limb \cref{fig:recon_nvae}.

\subsection{Direct Embeddings}
\begin{wrapfigure}{r}{0.5\textwidth}
  \vspace{-1em}
  \begin{center}
    \includegraphics[width=0.48\textwidth]{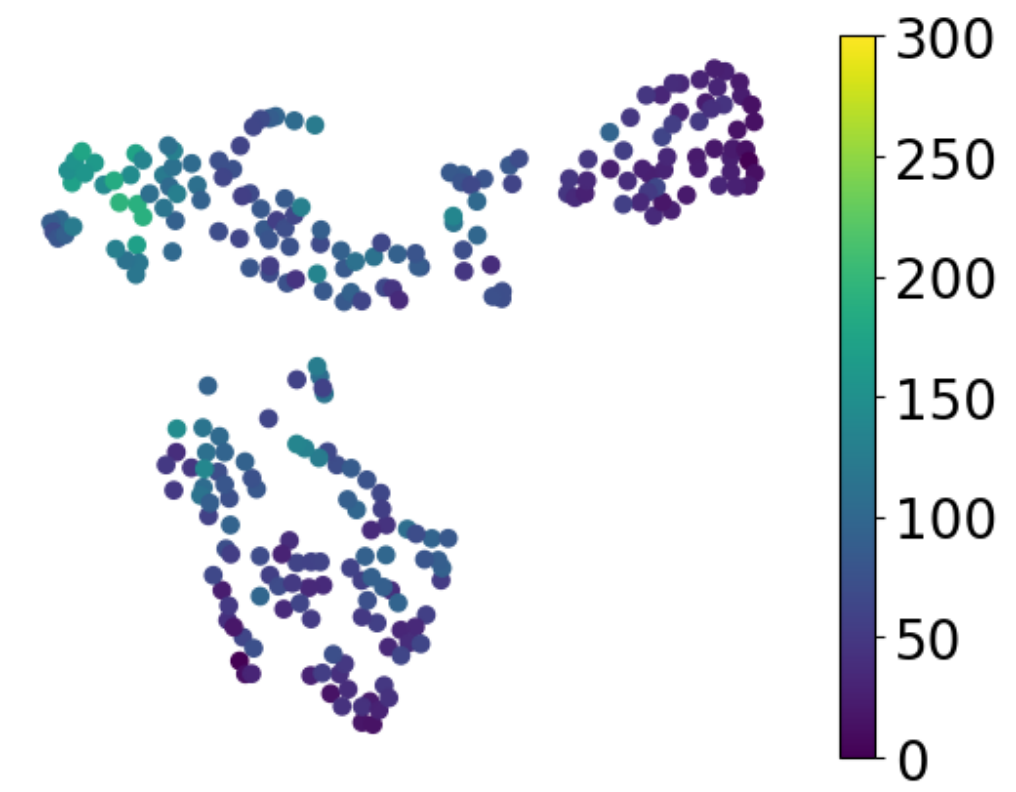}
  \end{center}
  \caption{TSNE applied to a small embedding sample, colored top left supporting to higher activity months.\label{fig:tsne}}
  \vspace{-1.25em}
\end{wrapfigure}

Training each scientific validation case on the embeddings directly led to generally much faster training time and matching or surpassing of accuracy. The was an effort made to best evaluate the embeddings outside of the scientific cases to consider embedding-to-embedding comparison. The common TSNE approach was taken over a small one-year sample, \cref{fig:tsne} indicates there was seperation by solar activity. This approach however is still fairly opaque and hence the validation approaches are considered more appropriate. 

\begin{figure}[h]
\includegraphics[width=0.8\textwidth]{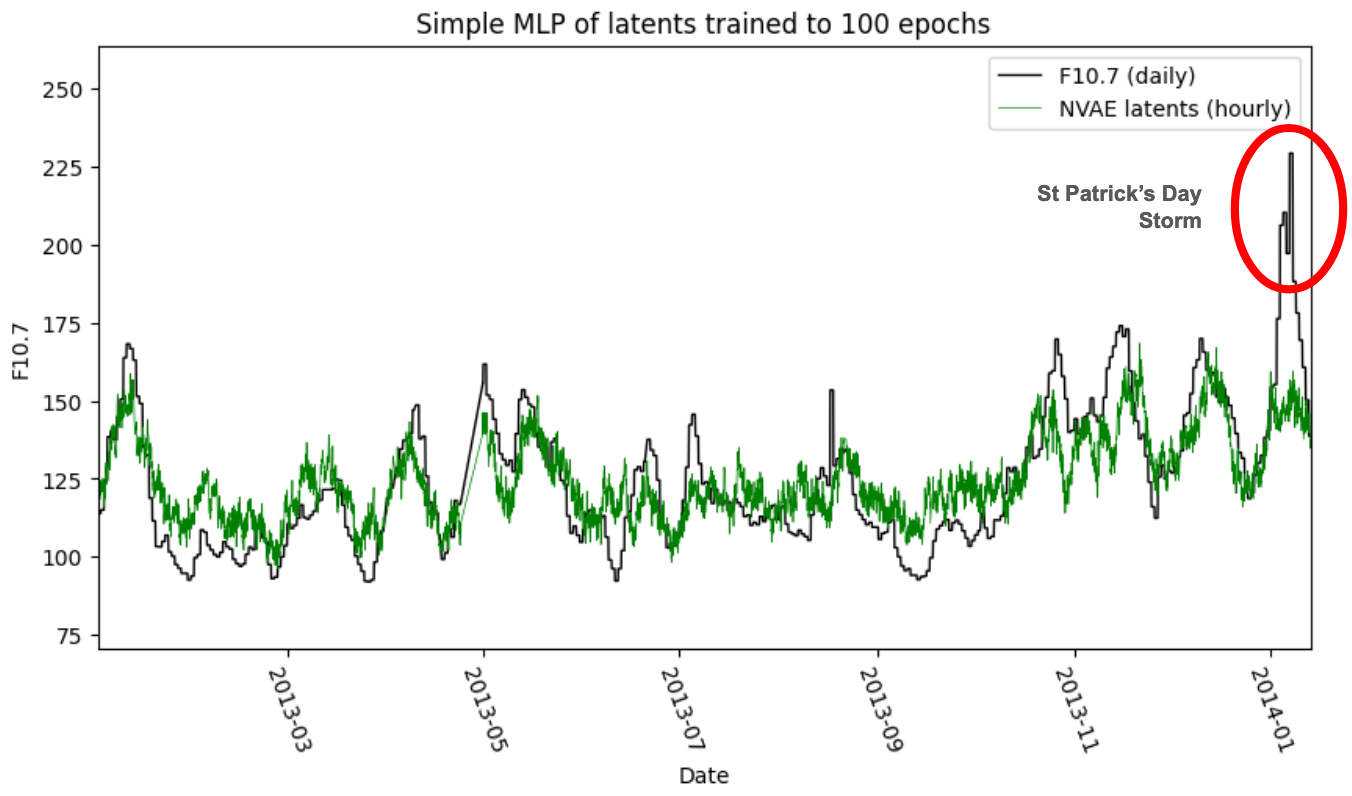}
\centering
\caption{Direct embedding prediction of F10.7 proxy.}
\end{figure}


\paragraph{Scaling} Scaling engineering effort on TPUs is a different set of skills than ML-engineering effort on GPUs. On GPUs the engineering allows for rapid model development and innovation; often possible on a local GPU machine. In contrast, TPUs are less accessible; requiring significant engineering to build a basic framework and often need to be reserved in advance through Google Cloud.  The decision between GPU and TPU would be easier if there was mature, hardware agnostic layers. Tools like PyTorch Lightning show promise in this area but still require significant engineering to develop.

\begin{figure}
    \centering
    \includegraphics[width=0.8\linewidth]{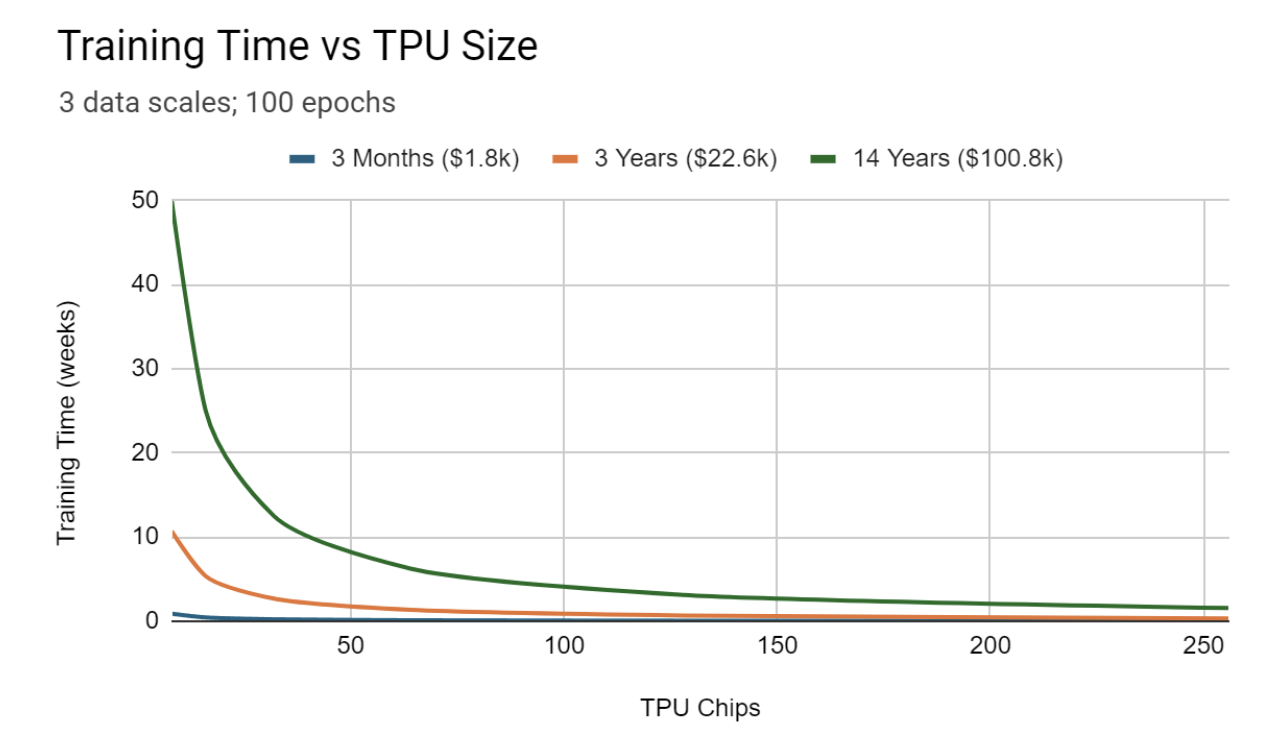}
    \caption{TPU scaling training time estimations over different sizes of TPU and total time period.}
    \label{fig:scaling}
\end{figure}

Finally, model training can be extrapolated roughly from historical training times. If we take a most closed-form scenario whereby one epoch takes 90 minutes to process 3 months of SDO AIA on a TPU with 16*8 = 128Gb available memory at \$1.5 per chip hour, scaling over various data training periods would look roughly like \cref{fig:scaling} and require on-disk requirements and VRAM to accommodate estimated in \cref{tab:disk_and_vram_req}.

\begin{table}[!h]
\begin{tabular}{@{}llllllll@{}}
\textbf{} & \multicolumn{2}{c}{\textbf{Precision}} & \multicolumn{2}{c}{\textbf{Size on disk (Gb)}} & \multicolumn{3}{c}{\textbf{Size on GPU (Gb)}} \\
\textbf{Parameter Size} & \textit{BF16} & \textit{32} & \textit{BF16} & \textit{32} & \textit{Quantized 8-bit} & \textit{BF16} & \textit{32} \\
\textit{330M} & 5B & 11B & 0.66 & 1.32 & 0.36 & 0.73 & 1.45 \\
\textit{700M} & 11B & 22B & 1.40 & 2.80 & 0.77 & 1.54 & 3.08 \\
\textit{2B} & 32B & 64B & 4.00 & 8.00 & 2.20 & 4.40 & 8.80 \\
\textit{10B} & 160B & 320B & 20.00 & 40.00 & 11.00 & 22.00 & 44.00
\end{tabular}
\caption{Size-on-disk and VRAM requirement estimates by increasing parameter size.\label{tab:disk_and_vram_req}}
\end{table}

\section{Discussion}

The adaptation of foundation models (FMs) to diverse scientific use-cases is a crucial step toward unlocking the full potential of data-driven research in space sciences. This initial exploration with the Solar Dynamics Observatory Foundation Model (SDO-FM) showcases the flexibility of these models within the context of space science, demonstrating that this new generation of large scale models can be successfully tailored to specific downstream objectives through two primary strategies: fine-tuning and leveraging latent representations. This research highlights the potential for reconfiguring embeddings to address a wide range of heliophysics tasks without the need for exhaustive retraining, simplifying the cost and complexity of using ML for heliophysics and space weather investigations. 

The first strategy: selective fine-tuning,  involves selectively freezing portions of the backbone model while allowing specific layers, often including the task-specific head, to continue learning. This selective fine-tuning allows for the preservation of learned representations in the backbone, while enabling specialization to new downstream tasks. By maintaining the core structure of the FM, we reduce the risk of catastrophic forgetting, a frequent challenge in transfer learning (where new learned representations are updated and optimized for the new task at the expense of prior knowledge) and optimize for both efficiency and precision. This approach is especially beneficial for scientific use-cases where domain-specific generalizations can be complex and retraining an entire model from scratch would be computationally prohibitive.

The second strategy: leveraging latent representations, offers computationally efficient approach by  bypassing the need for traditional fine-tuning. Instead, we extract compressed latent embeddings from the backbone, which capture high-level, abstract features of the input data. These latent embeddings can then serve as a basis for new models focused on downstream tasks such as reconstruction or prediction. This approach also drastically reduces the size of the SDO dataset, simplifying the input to subsequent models and handling complexity while maintaining the essential phenomenological information. Latent feature extraction is particularly valuable in environments where computational resources are limited; for example on a spacecraft; as it allows for faster training without adversely compromising outcomes. 

Both the selective fine-tuning and latent representation strategies were effective in adapting SDO-FM to downstream scientific tasks, marking a significant advancement in the development of foundation models using multi-modal space-acquired data. Our findings underscore the importance of structured governance in the ongoing development and deployment of foundation models for science, of which SDO-FM is an early exemplar. Foundation models are a new class of scientific tool and, in line with the principles guiding traditional scientific instruments, we propose that all science-based foundation models should be overseen by a principal investigator (PI), supported by a multidisciplinary working group. This group would include scientists, data providers, and data scientists, ensuring broad awareness of the model's capabilities, effective onboarding, and clear communication regarding versioning and expectations. For instance, this initial version of SDO-FM, with its relatively low temporal resolution can only emulate proven scientific results. Thus, it is essential for future versions to have clearly defined science goals and user expectations, where the science requirements are driving the engineering choices.  

Adherence to open science and transparency around model uncertainty should be central to continued development of foundation models. Comprehensive validation protocols, such as those established by Munoz-Jaramillo’s community-built \href{https://github.com/SwRI-IDEA-Lab/sw-forecast-protocol}{framework for forecasting} should be integral to the model’s development, analogous to aerospace industry flight checks, particularly in cases where there may be imbalances in the data. Moreover, benchmarking SDO-FM against other foundation models and assessing its performance across both in-distribution and out-of-distribution scenarios will be critical for ensuring its scientific robustness. Shared definitions of progress, coupled with transparent publication of failure modes in accordance with NASA's \href{https://nasa.github.io/Transform-to-Open-Science/}{Transform to Open Science} initiative, will facilitate ongoing improvements, accountability, and trust within the scientific community and ensure this shared resource is used. Ultimately, the development of a robust, adaptable foundation model in the field of space science hinges on a coordinated effort that integrates scientific rigor, engineering best practices, and an unwavering commitment to transparency and accessibility.

\begin{ack}

This work is the research product of the SDO-FM: A Multi-Modal Foundation Model POC for SDO (Grant\#: 80NSSC24K0701); funded and supported by NASA. The research and its outputs have been designed, managed, and delivered by Trillium Technologies Inc.

\end{ack}


\bibliography{refs}





\end{document}